# Research into the sampling methods of digital beam position measurement

WU Weihao (邬维浩)[1], ZHAO Lei (赵雷)[1,*], CHEN Erlei (陈二雷)[1], LIU Shubin (刘树彬)[1], AN Qi (安琪)[1]

[1]*State Key Laboratory of Particle Detection and Electronics，USTC，Hefei，230026*

*E-mail: zlei@ustc.edu.cn*

**Abstract:** BPM (Beam Position Measurement) system is one of the most important beam diagnostic instruments in accelerators. A fully digital BPM (DBPM) has been designed for SSRF (Shanghai Synchrotron Radiation Facility). As Analog-to-Digital Converter (ADC) is one crucial part in the DBPM system, the sampling methods should be studied to achieve optimum performance. We implemented different sampling modes and compared them through tests. Meanwhile, long term variation among four sampling channels is another concern, which would introduce errors in beam position measurement. We designed an interleaved distribution scheme to address this issue. To evaluate these sampling methods, we conducted commissioning tests with the beam in SSRF. Test results indicate that with proper sampling methods, a turn-by-turn (TBT) position resolution better than 1 μm is achieved, and the slow-acquisition (SA) position resolution is enhanced from 4.28 μm to 0.17 μm successfully with interleaved distribution applied, both beyond requirement.

## 1. Introduction

SSRF is one of the third generation synchrotron radiation sources in the world, and it has become an important platform in various scientific research domains [1]. In SSRF, BPMs are indispensable beam diagnostic systems that can measure the transverse

---

This work was supported by the Knowledge Innovation Program of the Chinese Academy of Sciences (KJCX2-YW-N27) and the National Natural Science Foundation of China (No.11205153 and 11175176).

* Corresponding author. Tel.: 0551-63607746. E-mail address: zlei@ustc.edu.cn

centroid of the beam [2-6]. A DBPM system has been designed for SSRF [5, 6], the block diagram of which is shown in Fig. 1. The DBPM can be used in many machines, like a linear accelerator (LINAC), a LINAC to booster transfer line (LTB), a full energy booster (BS), a booster to storage ring transfer line (BTS) and a storage ring (SR) [7]. It receives signals from four capacitive pickup electrodes positioned around the beam pipe. The buttons generate fast pulses at the electron bunch repetition rate of $f_{RF}$=499.654 MHz [1]. The input signals are first conditioned by the analogue Radio Frequency (RF) circuits in four sampling channels of the DBPM, with band-pass filters and amplification circuits. Using RF amplifiers and attenuators, a dynamic range of more than 50 dB is achieved [6]. Then, the 499.654 MHz RF signals are digitized by four 14-bit ADCs with a sampling rate of around 117.28 MHz, through which digital Intermediate Frequency (IF) signals of around 30.5 MHz are generated. The digital IF signals are further processed by the algorithms integrated within one FPGA (Field-Programmable Gate Array) device.

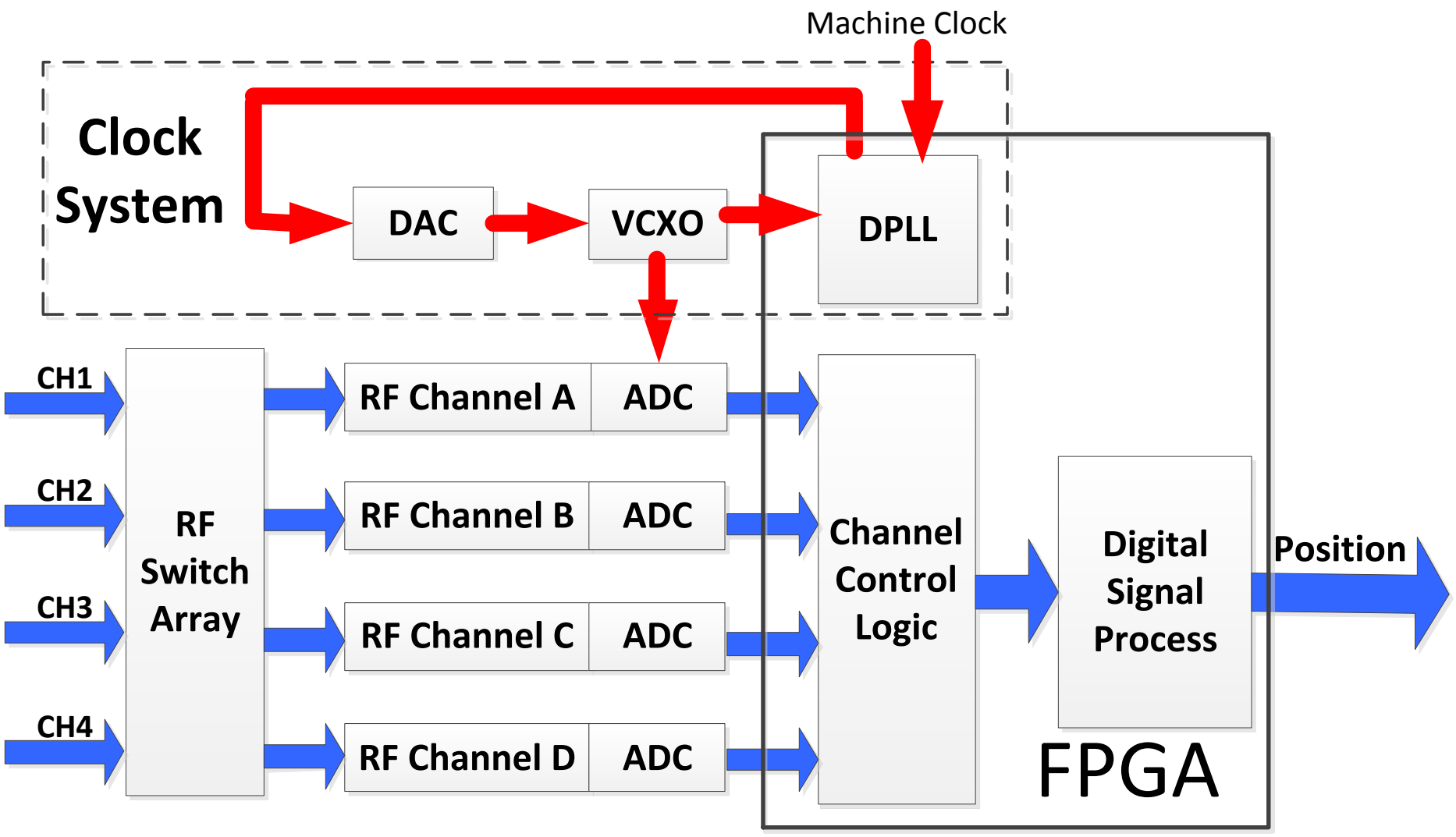

**Fig.1**. Block diagram of the DBPM designed for SSRF.

As shown in Fig. 2, the IF signals are converted into I and Q arrays by Digital Down Converters (DDCs) before the amplitudes of four input signals can be calculated by the Cordic algorithm. Then by employing the $\Delta/\Sigma$ principle [8], the beam position can be calculated using Eq. (1) and (2) as follows

$$X = K_x \times \frac{V_1 - V_2 + V_3 - V_4}{V_1 + V_2 + V_3 + V_4}, \qquad (1)$$

$$Y = K_y \times \frac{V_1 + V_2 - V_3 - V_4}{V_1 + V_2 + V_3 + V_4}, \qquad (2)$$

where $K_x$, $K_y = 10\ \text{cm}$ is the effective length factors in X and Y directions.

At the SSRF the harmonic number of the ring is $H$ = 720, so the turn rate is given by the machine clock frequency ($f_{mc}$) 499.654 MHz / $H$ = 693.964 kHz. Using a Low-Pass-Filter (LPF), the SA data can be obtained from TBT data. While the TBT data is used to analyze the spectrum of the beam position and study the position noise between tens of kHz to hundreds of kHz, the SA data is applied to beam monitoring, as well as in a feedback loop to stabilise the beam orbit [7].

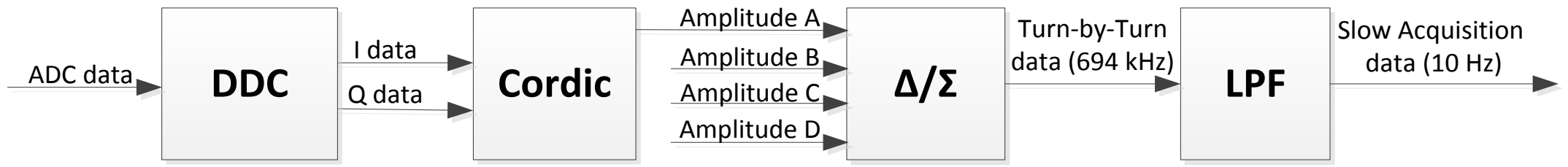

**Fig.2**. Block diagram of the digital signal processing.

The performance of the ADCs has direct influence on the measurement resolution of the overall BPM system; therefore, great care must be taken with which sampling method is used. In this paper, we explored the synchronization and an off-tune sampling modes, implemented them in electronics design and evaluated the performance through tests. A Digital Phase-Locked Loop (DPLL) algorithm was deployed on an FPGA device, which enables the system to implement and switch between the different sampling methods used by the ADCs. To address the issue of long-term variation of the ADCs and the RF circuits, a switch array was designed to implement interleaved distribution of the input signals among four sampling channels. Finally, the commissioning tests results of different sampling methods in the SSRF are introduced and compared, followed by the conclusion.

## 2. Sampling methods in the ADCs

The electron bunches produce RF signals in the BPM pick-ups with a frequency of $f_{RF}$ = 499.654 MHz, which are fed into the DBPM inputs. These signals are modulated by a macro pulse due to the turn frequency given by $f_{mc}$ = 693.964 kHz ($f_{RF}$ / $H$). If the ADC sampling clock was not synchronized to the machine clock ($f_{mc}$), there would be a variation in the number of RF signals per TBT cycle [6]. This variation would cause undue fluctuations in the amplitude measurements, which could finally deteriorate position resolution. Based on the above consideration, the synchronization sampling becomes mandatory for the TBT data which means the sampling frequency ($f_s$) should be an integer multiple of $f_{mc}$, as shown in Eq. (3). In such a case, the TBT data rate is exactly equivalent to $f_{mc}$.

$$\frac{f_{in}}{f_s} = \frac{720 \times f_{mc}}{169 \times f_{mc}} = \frac{720}{169}. \qquad (3)$$

Regarding the SA data, the requirement on the sampling frequency is quite different. As mentioned previously, the TBT data (694 kHz) is filtered by a digital LPF to obtain the SA data with a much lower rate of around 10 Hz; therefore, even if the sampling clock was not synchronized to the machine clock, the amplitude fluctuation could be filtered out. On the other hand, due to the modulation of the macro pulse, sidebands with a frequency interval of $f_{mc}$ exist in the frequency spectrum of the digital IF signal. In the ADCs, the non-linearity of the circuits would cause harmonics of the digital IF signal to appear, which is accompanied by a similar distribution of sidebands. Some of the sidebands around the harmonics would overlap at the digital IF signal frequency ($44 \times f_{mc}$), which would cause measurement errors with a very low frequency. This has been observed in the SA measurement results of some BPM instrumentations, such as Libera Electron [9]. To address this issue, we should shift $f_s$ in Eq. (3) by a certain small frequency difference $\Delta f$ to relocate the sideband of the harmonics outside of the pass-band (5 Hz) of the LPF for the SA data [9]. A straightforward idea is to set $f_s$ totally independent from $f_{mc}$, but the problem is that $\Delta f$ cannot be controlled precisely as $f_{mc}$ shifts along with time. In the research of this paper, we applied an off-tune sampling mode, in which $\Delta f$ is set to a certain value that can also be modified easily by commands. To accommodate the above sampling modes and switch between them, a DPLL is specially designed based on an FPGA device.

## 3. Synchronization and Off-tune Sampling based on DPLL

In conventional methods for clock synchronization, commercially available PLL chips are employed, which consist of a phase detector, a charge pumper, and a loop filter implemented in ICs. This is an easy way to design sampling clock generation circuits, however, rendering it impossible to switch between different sampling modes. To achieve good flexibility, we adopted a digital solution based on a modified DPLL in an FPGA device (XC4VFX100-11ff1152 from Xilinx virtex-4 family). The output codes from the DPLL are used to control an external digital-to-analog convertor (DAC) whose output is further fed to a voltage-controlled crystal oscillator (VCXO), as shown in Fig. 3. With different division factors ($M$ and $N$ in Fig. 3), the frequency relationship between the output clock ($f_s$) and the input reference clock ($f_{mc}$) can be determined.

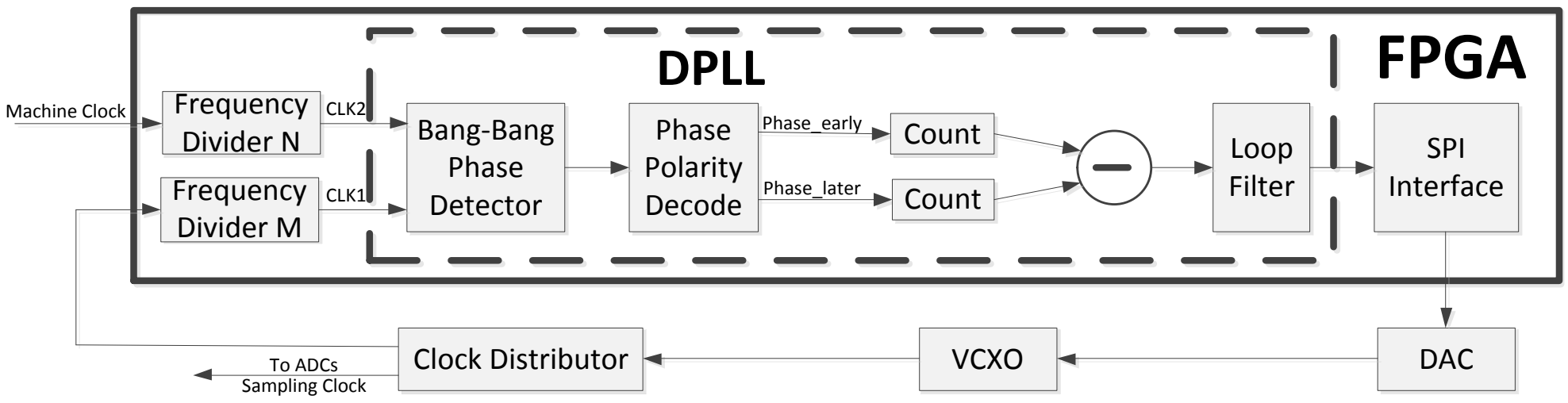

**Fig.3**. Block diagram of the clock generation circuits.

As one of the kernel parts in DPLL, the digital phase detector is categorized into the linear type like Hogge phase detector and the nonlinear type like Alexander phase detector (i.e. Bang-Bang phase detector) [10] [11]. Considering the superior performance and simplicity of the Alexander phase detector, it is chosen in the DBPM system [12]. The Alexander phase detector outputs 3-bit data indicating which one of the two input signals leads the other. Fig. 4 (a) and (b) shows its structure and principle, respectively; *A*, *T* and *B* are three samples taken by three consecutive clock edges. If “CLK2” leads “CLK1”, the first sample (*A*) is unequal to the last two (*T* and *B*).

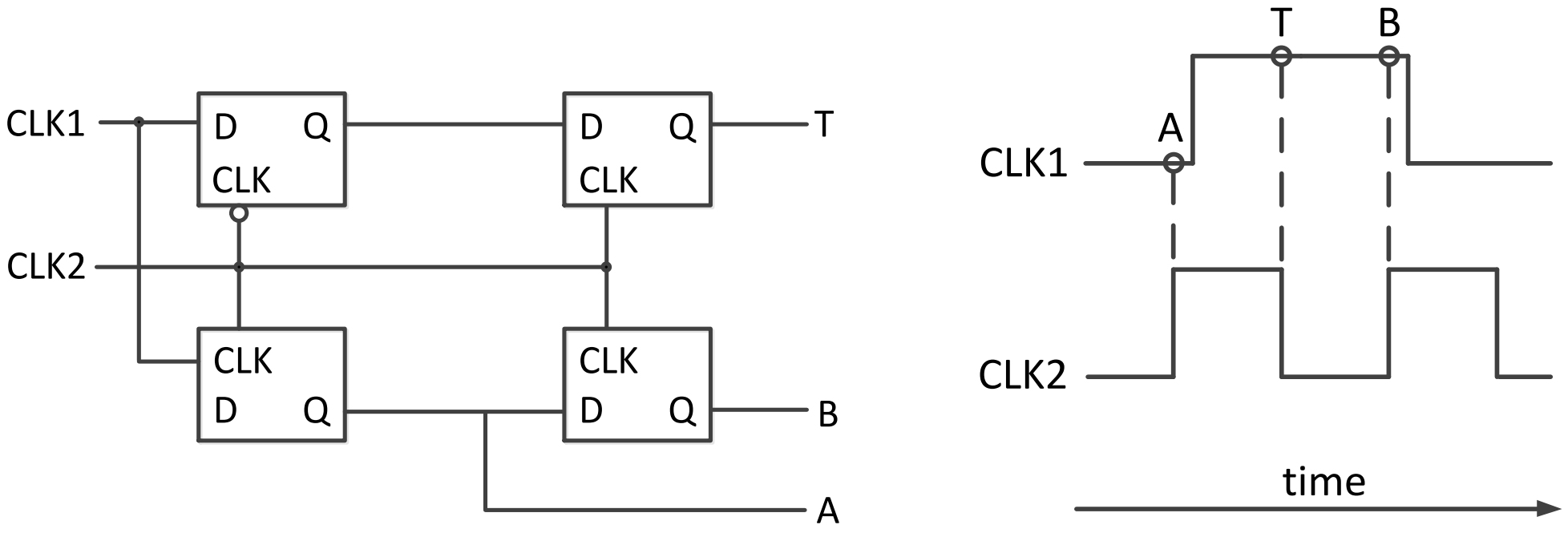

**Fig.4**. (a) structure of Bang-Bang phase detector (a); (b) principle of Bang-Bang phase detector.

We implemented this digital phase detector in the FPGA, and conducted simulations based on ISim (ISE 14.4). The result concords well with the theoretical analysis, as shown in Fig. 5. Connecting this phase detector with the loop filter and other components in Fig. 3 implements a whole DPLL. With this scheme, there exists a relationship between the frequencies of two input clocks of the DPLL when in the locked state, as shown in

$$f_{clk2} = 2 \times f_{clk1.} \qquad (4)$$

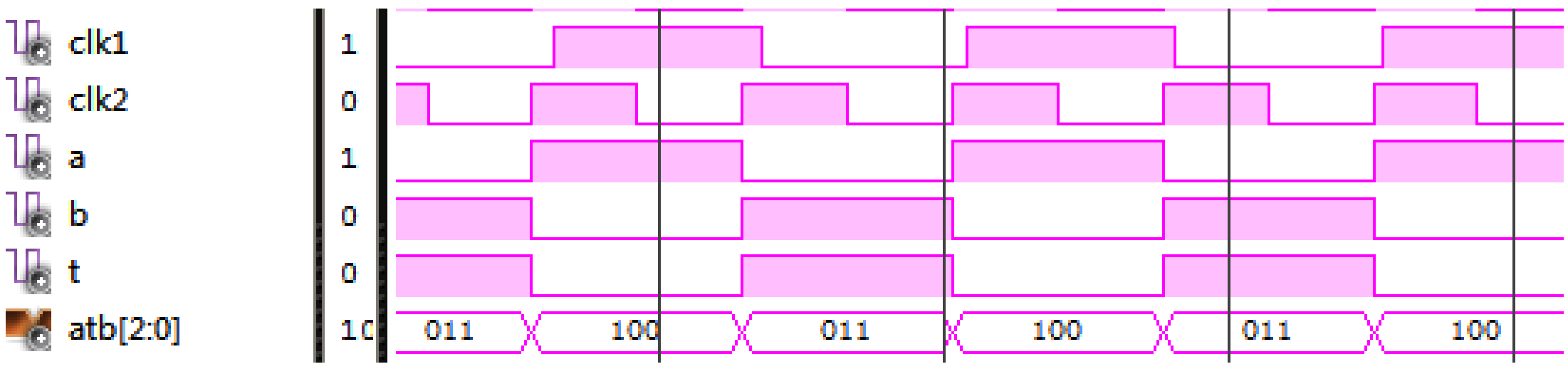

**Fig.5**. Simulation result of Bang-Bang phase detector.

Considering the relationship among the machine clock, the ADC sampling clock, CLK1and CLK2 in Fig. 3, $f_s$ can expressed as in Eq. (5). Then the relationship in Eq. (3) can be achieved with $M$=338 and $N$=1.

$$f_s = \frac{M}{2N} \times f_{mc}. \quad (5)$$

Meanwhile, as mentioned above, to prevent the sidebands of the harmonics from deteriorating the SA data, $f_s$ should be shifted by a certain frequency difference $\Delta f$ (i.e. the off-tune sampling mode). In this design, we implemented $\Delta f$ by configuring a frequency divider with a factor $K$, as shown in

$$\Delta f = \frac{f_{mc}}{K}. \quad (6)$$

Thus, in the off-tune sampling mode, $f_s$ can be now expressed as

$$f_s = 169 \times f_{mc} + \Delta f = 169 \times f_{mc} + \frac{f_{mc}}{K} = \frac{169 \times K + 1}{K} \times f_{mc}. \quad (7)$$

Comparing with Eqs. (5) and (7), the off-tune sampling can be implemented by the DPLL, and the relationship among $N$, $M$ and $K$ is shown in

$$N = \frac{K}{2}; \quad M = 169 \times K + 1. \quad (8)$$

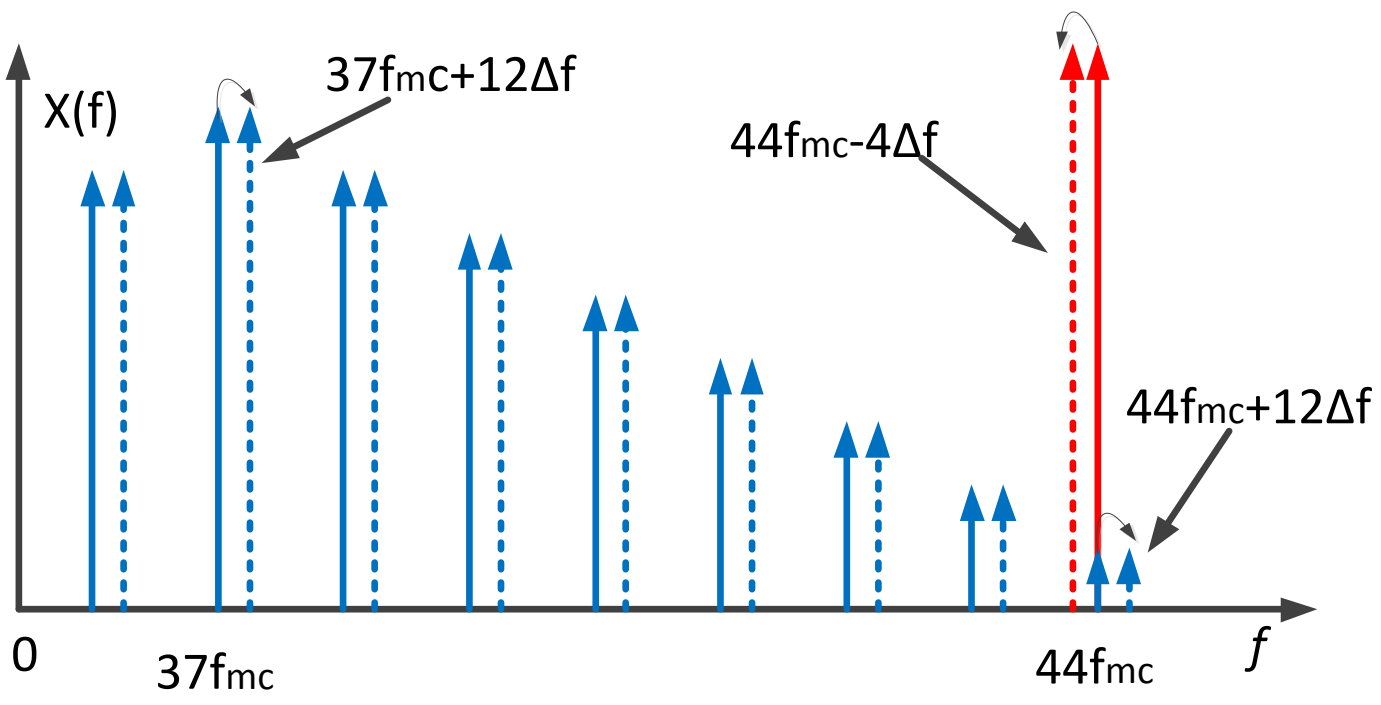

**Fig.6**. Sidebands of 3rd harmonic and IF baseband signal.

The next step is to determine the proper value of $\Delta f$. The major influence comes from the sideband ($44 \times f_{mc}$) of the third harmonic ($37 \times f_{mc}$) overlapping on the digital IF signal [9]. As shown in Fig. 6, in the off-tune sampling mode, the sideband of the third harmonic and the digital IF signal are shifted by $12 \times \Delta f$ and $-4 \times \Delta f$, respectively, i.e. a total frequency interval of $16 \times \Delta f$ is achieved. To guarantee that the sideband can be effectively suppressed by the LPF (5 Hz pass band) for the SA data, the value of $\Delta f$ should meet the requirement as presented in

$$16 \times \Delta f > 5\, Hz. \quad (9)$$

Meanwhile, considering the frequency tuning range of the VCXO (VFVX 120), the upper limitation of $\Delta f$ is

$$\Delta f < 5.85\, kHz. \qquad (10)$$

Combining Eqs. (9), (10) and (6), we can obtain the proper value of $K$, which ranges from 118 to 2221.

$$\frac{5}{16}\, Hz < \frac{f_{mc}}{K} < 5.85\, kHz \qquad => 118 < K < 2221. \qquad (11)$$

In the design of DPLL, $K$ is chosen to be 256, and the corresponding values of $N$ and $M$ are 128 and 43265. The final sampling frequency in the off-tune mode can be expressed in

$$f_s = \frac{M}{2N} \times f_{mc} = \frac{43265}{2 \times 128} f_{mc} = 169 \times f_{mc} + \frac{1}{256} \times f_{mc}. \qquad (12)$$

## 4. Consideration of Long-term Stability

In the DBPM, the SA data are mainly used to monitor long term variation of the beam position [7]. The beam position is calculated from the input signal amplitudes $(A_1, A_2, A_3, A_4)$ multiplied by the gains of four sampling channels $(G_1, G_2, G_3, G_4)$, as shown in

$$X = K_x * \frac{V_1 - V_2 + V_3 - V_4}{V_1 + V_2 + V_3 + V_4} = K_x * \frac{G_1 * A_1 - G_2 * A_2 + G_3 * A_3 - G_4 * A_4}{G_1 * A_1 + G_2 * A_2 + G_3 * A_3 + G_4 * A_4}, \quad (13)$$

$$Y = K_y * \frac{V_1 + V_2 - V_3 - V_4}{V_1 + V_2 + V_3 + V_4} = K_y * \frac{G_1 * A_1 + G_2 * A_2 - G_3 * A_3 - G_4 * A_4}{G_1 * A_1 + G_2 * A_2 + G_3 * A_3 + G_4 * A_4}, \quad (14)$$

where $K_x$, $K_y = 10$ cm is the effective length factors in X and Y directions; $G_n$ is the gain of the sampling channel; $A_n$ is the amplitude of the input signal.

In the ideal situation, the four gain factors are equal and there are no measurement errors in beam position results; however, it is not the case in real applications. If the difference among the four gain factors remain constant, the measurement error in SA data can easily be corrected by off-line calibration. However, we found that these gain factors change quite differently which would inevitably deteriorate the resolution of the SA data in long term. As shown in Fig. 7, obvious fluctuation of the SA data can be observed (the corresponding position resolution is 0.832 μm) in the laboratory tests where the input signals are generated by a signal sources (ROHDE & SCHWARZ SMA 100A). Therefore, we must apply new methods to address this issue.

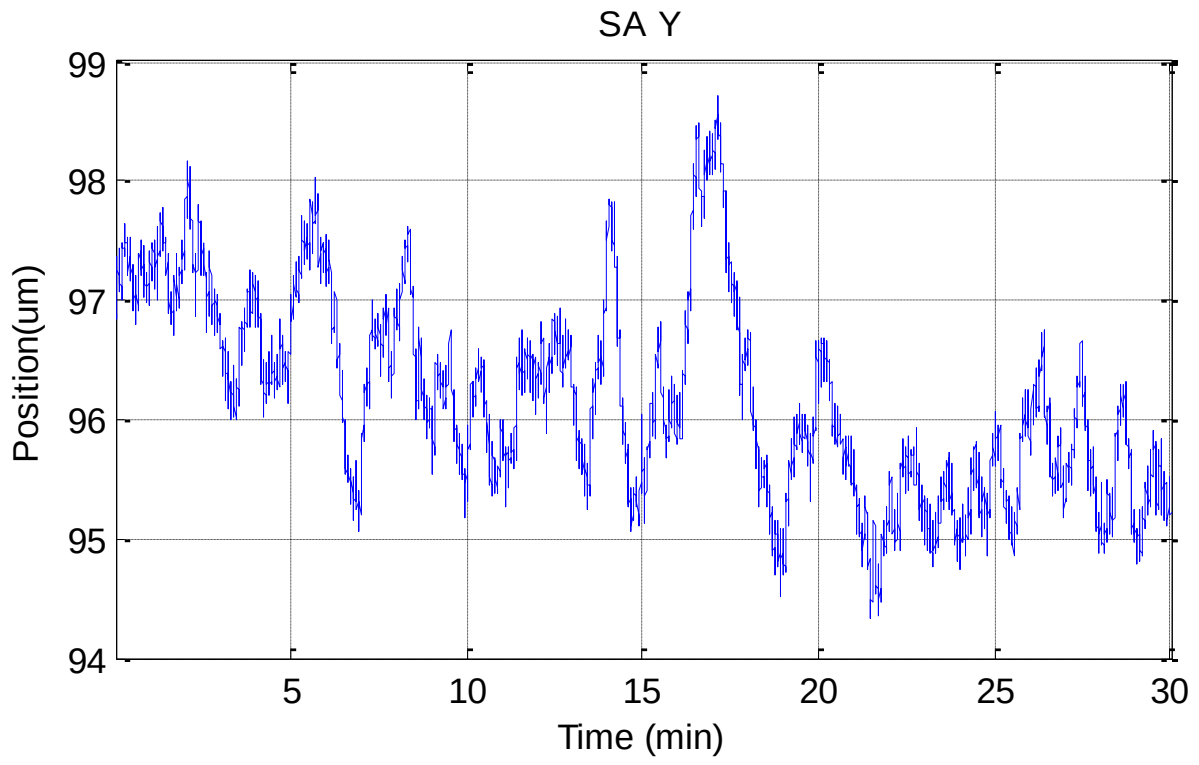

**Fig.7**. Y position waveform of the SA data.

The basic idea is to distribute the four input signals to the four sampling channels alternately according to a specified sequence. We designed an interleaved distribution scheme by using an RF switch array whose structure is shown in Fig. 8 (a) [13]. With different controlling signals, this switch array establishes corresponding signal paths between the four input ports and the four sampling channels. The controlling signals are generated by the FPGA, and are organized in a special sequence, with which the four input signals are evenly interleaved to 4 channels, as shown in Fig. 8(b). The switching frequency is designed to be around 5.42 kHz which is much larger than the pass band (5 Hz) of the LPF for the SA data in Fig. 2. So the distortion introduced by the switching process can easily be filtered out. In this case, the four sampling channels have equivalent influence on the processing of each input signal, which means the equality of effective gains for all input signals, i.e. the measurement errors caused by the long term fluctuation of electronics are greatly suppressed.

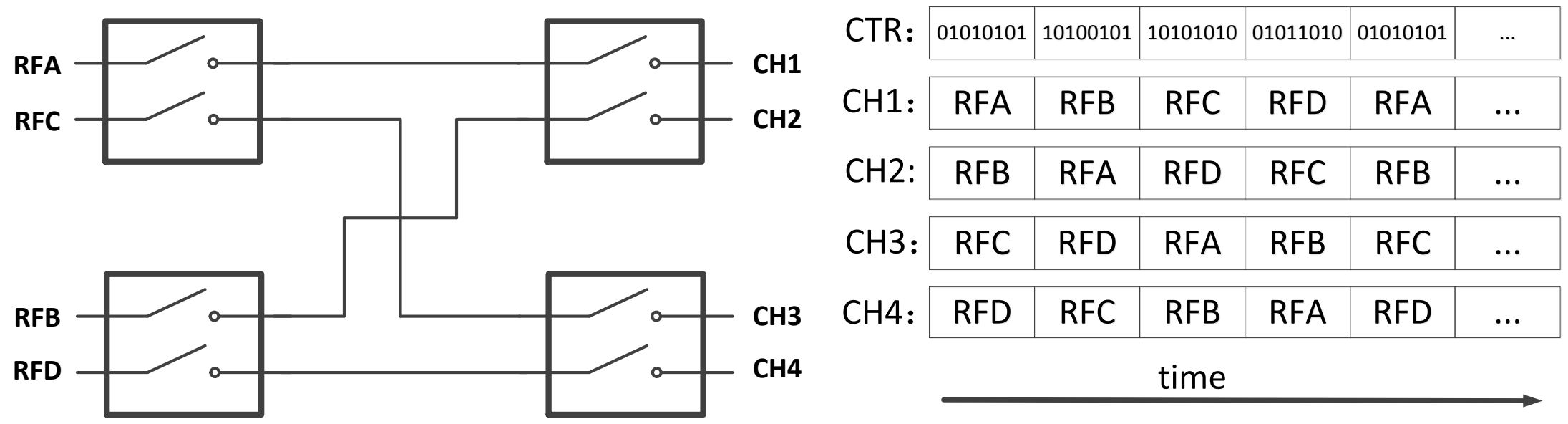

**Fig.8**. (a) Structure of the switch array; (b) time diagram of the switch array.

Initial tests in the laboratory have been conducted and the result is shown in Fig. 9. We can find that the position information of the SA data in thirty minutes is more stable and the resolution is enhanced from 0.832 μm to 0.032 μm, compared with the result in Fig. 7.

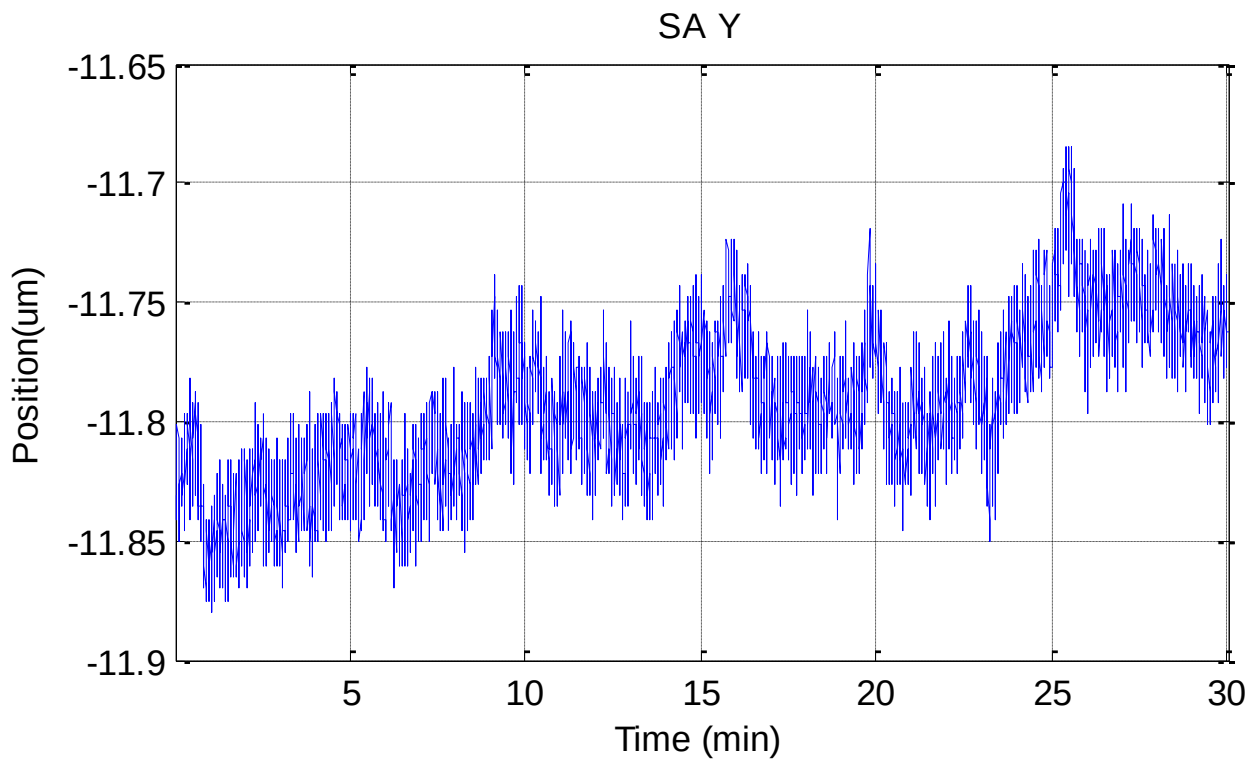

**Fig.9**. Y position waveform of the SA data with the interleaved distribution scheme.

## 5. Test Results

To evaluate the overall performance of the DBPM in different sampling methods, we conducted commissioning tests with the SSRF accelerator, as shown in Fig. 10. The modulated RF signals from pickups were sent to the DBPM through four coaxial cables. After manipulating the signals, the DBPM sent the position information of beam in the accelerator to a remote PC through ethernet for further analysis.

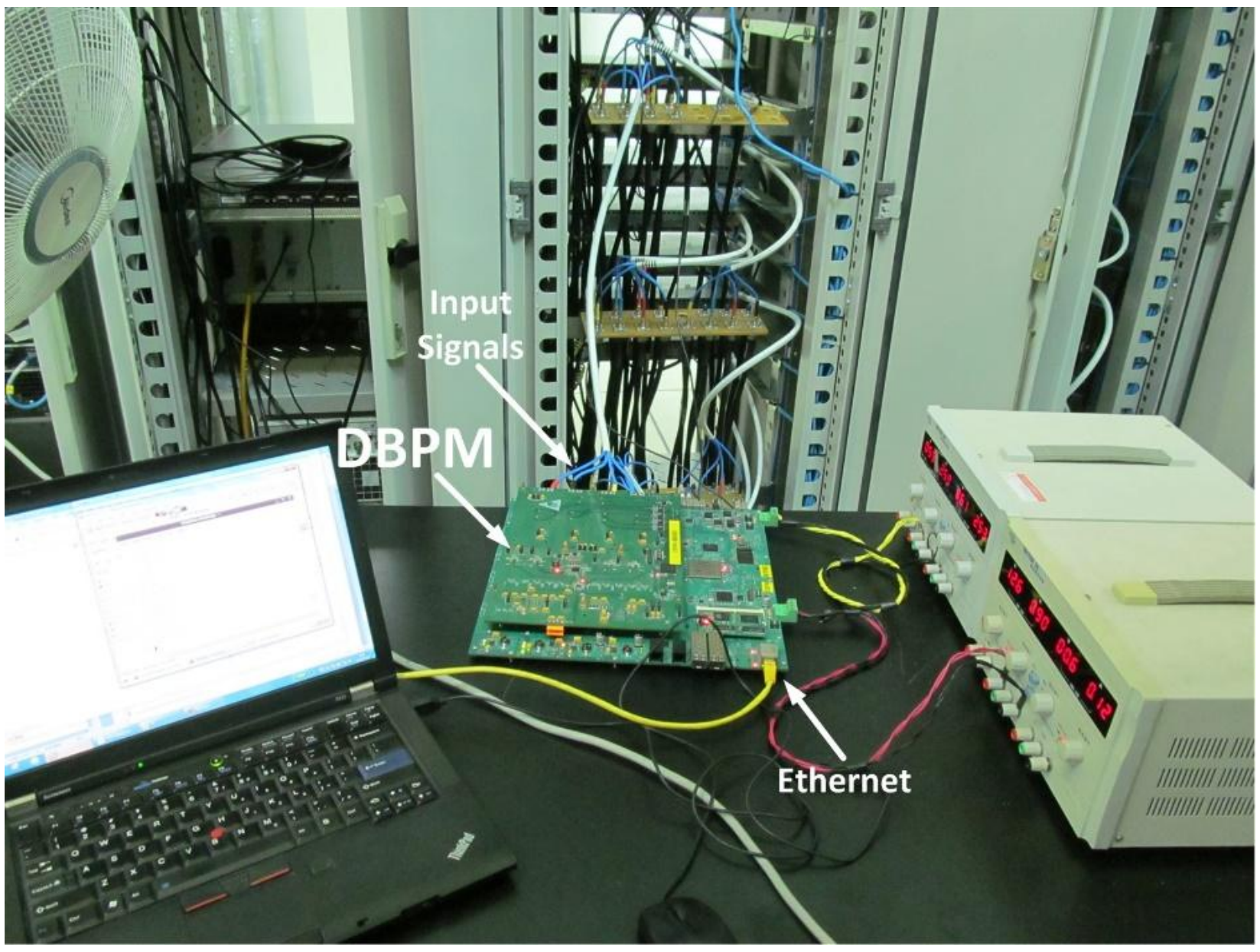

**Fig.10**. DBPM under commissioning tests in SSRF.

## 5.1 Results of the TBT data

We tested the TBT data in two modes: the synchronization sampling and the off-tune sampling. The normalized signal amplitudes of the TBT data in both sampling modes are shown in Fig. 11. Fluctuation can be clearly observed in Fig. 11 (b), just as discussed in Section 2. Therefore, the synchronization sampling is chosen for the TBT data, and the position resolution is around 0.61 μm.

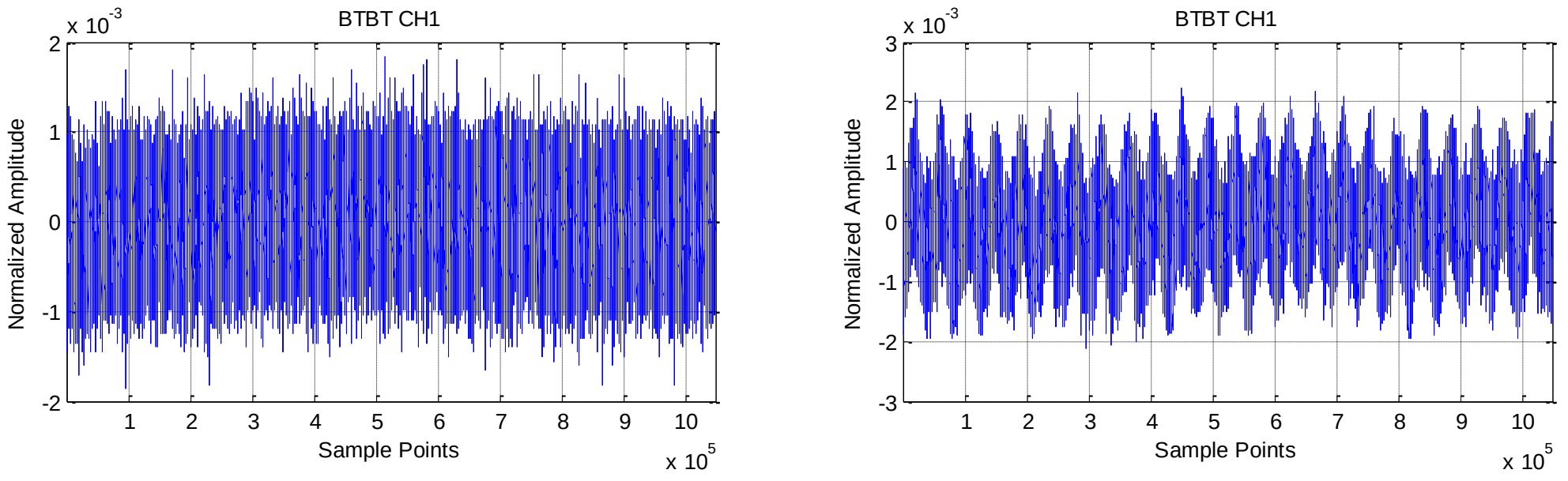

**Fig.11**. (a) Amplitude in the synchronization sampling; (b) Amplitude in the off-tune sampling

## 5.2 Results of the SA data

As presented in Fig. 12, the interleaved distribution scheme that we designed can increase the performance of the SA data significantly; the position resolution of a thirty minute data collection test is enhanced from 4.28 μm to 0.17 μm. And similar performance can be achieved in the synchronization sampling and the off-tune sampling, as shown in Fig. 12 (b) and (c). This is probably due to the good performance of the analog frond-end in the DBPM system.

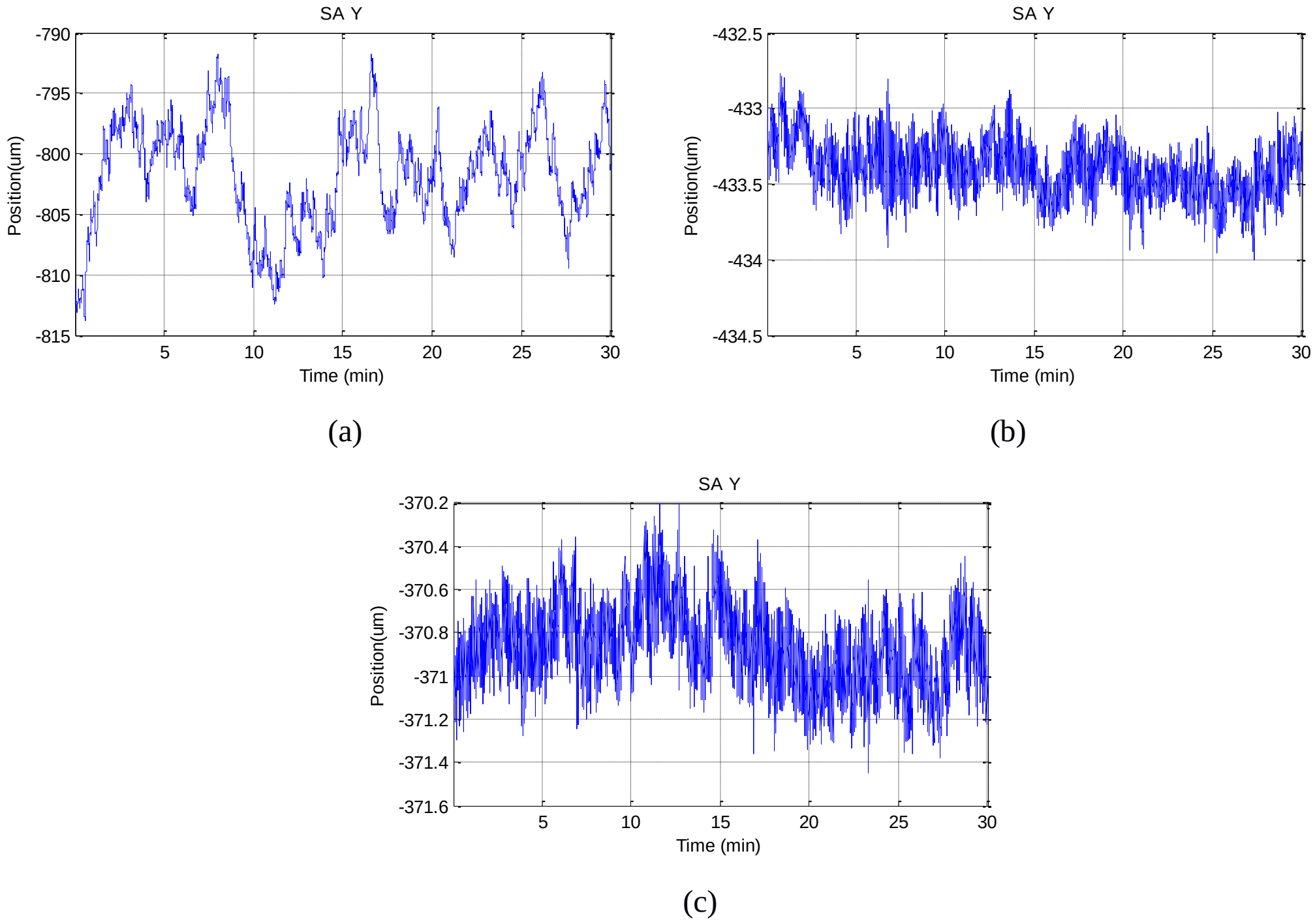

**Fig.12**. Y position waveform of the SA data

(a) Without interleaved distribution (rms=4.28 μm)

(b) With interleaved distribution and the synchronization sampling (rms=0.16 μm)

(c) With interleaved distribution and the off-tune sampling (rms=0.17 μm)

## Conclusion

Different kinds of sampling methods are studied to optimize the performance of the DBPM designed for the SSRF, including the synchronization sampling, the off-tune sampling and the interleaved distribution scheme which aims to address the long term stability. The commissioning test results with the SSRF accelerator indicate that with an optimized sampling method, a TBT position resolution better that 1 μm is achieved. It was found that the SA resolution can be effectively enhanced to 0.17 μm using an interleaved distribution scheme. Both the position resolutions of the TBT and the SA data exceed the requirements.

## Acknowledgment

The authors would like to thank all of the SSRF collaborators who helped this paper possible.